\documentclass[12pt]{article}
\usepackage{amsmath}
\usepackage{amssymb}

\newcommand{\be}{\begin{equation}}
\newcommand{\bea}{\begin{eqnarray}}
\newcommand{\ee}{\end{equation}}
\newcommand{\eea}{\end{eqnarray}}

\def\theequation{\arabic{section}.\arabic{equation}}
\begin{document}
\topmargin -1cm \oddsidemargin=0.25cm\evensidemargin=0.25cm
\setcounter{page}0
\renewcommand{\thefootnote}{\fnsymbol{footnote}}
\begin{titlepage}
\begin{flushright}
DFTT-8/2007\\
\end{flushright}
\vskip .7in
\begin{center}
{\Large \bf Interacting Higher Spins and the High Energy Limit of
the Bosonic String } \vskip .7in {\large Angelos
Fotopoulos$^a$\footnote{e-mail: {\tt foto@to.infn.it}} and Mirian
Tsulaia$^{b}$\footnote{e-mail: {\tt tsulaia@physics.uoc.gr}}} \vskip
.2in {$^a$ \it Dipartimento di Fisica Teorica dell'Universit\`a di
Torino and INFN Sezione di Torino via
P.Giuria 1, I-10125 Torino, Italy} \\
\vskip .2in { $^b$ \it Department of Physics and Institute of Plasma
Physics, University of Crete, 71003 Heraklion, Crete, Greece }
\\

\begin{abstract}
 In this note, we construct a BRST
invariant cubic vertex for  massless fields of arbitrary mixed
symmetry in flat space-time. The construction is based on the vertex
given in bosonic Open String Field Theory.
 The  algebra of gauge transformations is closed without any
additional, higher than cubic, couplings due to the presence  of an
infinite tower of massless fields. We briefly discuss the
generalization of this result to a curved space-time and other
possible implications.

\end{abstract}

\end{center}

\vfill

\end{titlepage}

\tableofcontents

\section{Introduction}

While the low energy limit of string theory is fairly well
understood the properties of its high energy limit
\cite{Gross:1987ar}-- \cite{Moeller:2005ez} (see also
\cite{Bandos:2007qn})
 still impose a series of questions. Several proposals and conjectures have been made
about the high energy limit of String Theory and its role in the
still mysterious M--theory.

For example, unlike various types of supergravities which describe
the low energy approximation of superstring theories, the field
theory description of the high energy regime of string theory is
still unknown. It is also not known which kind of gravitational
background is required for its consistency. Obviously, when dealing
with very small distances, the only type of nontrivial background a
string still might "feel" is a highly curved background.

Several results based on AdS/CFT duality strongly support the idea
that there might be a relation between the  high energy limit of
String theory and Higher Spin (HS) gauge theories
\cite{Sezgin:2002rt}. These and other considerations suggest that a
possible candidate (if any) for a field theoretical  description of
String theory at high energies might be the HS gauge theory
developed in \cite{Vasiliev:1990en}. This theory is defined on an
AdS background in any space-time dimension and classically  is a
perfectly consistent theory. One can wonder if there is any precise
connection between HS theory and string theory. For this reason it
seems interesting to formulate HS field theory in a language similar
to that used in String Theory like for example bosonic String Field
Theory (SFT) \cite{Francia:2002pt, Sagnotti:2003qa, Francia:2006hp}.
Moreover one can try to find a generalization of the methods used in
SFT to the case of curved backgrounds with constant curvature, such
as AdS space-times
\cite{Fotopoulos:2006ci}--\cite{Buchbinder:2006eq}.

A very useful method for describing HS theories is the BRST method
developed for example in \cite{Fotopoulos:2006ci, Buchbinder:2001bs,
Buchbinder:2007ak, Buchbinder:2006ge} and the   free theory
formulation of HS theories  based on the BRST approach has been
explored to some extent. The problem of constructing interaction
vertices at the high energy limit of String theory was considered
previously by various authors \cite{Koh:1986vg,Bonelli:2003kh} or alternatively one
can address the problem of an interaction between higher spin fields
without any recourse  to string theory \cite{Vasiliev:1990en,
Bekaert:2005jf}.

Since the  BRST description for various massive and massless fields
belonging to {\it irreducible} representations of Poincare and AdS
groups is available (see for example \cite{Fotopoulos:2006ci,
Buchbinder:2001bs, Buchbinder:2007ak, Buchbinder:2006ge}),  one can
try to find (self)interacting Lagrangians for them.  Here though, we
present another SFT inspired solution which proves to be exact to
all orders of the coupling constant. The interaction we consider
couples an infinite tower of {\it reducible} representations of the
Poincare group, rather then irreducible ones. The reason behind this
is that the problem of finding an interaction between reducible
representations shares a lot of similarity with String Theory and
this similarity might prove to be very helpful
\cite{Sagnotti:2003qa}. Moreover finding a formal analogy with
string theory one can hope to understand a ''physical" connection
between these theories.

The cubic interaction discussed in the present paper is obtained
from a formal high energy limit of conventional bosonic Open
String Field Theory (OSFT) \cite{Witten:1985cc, Gross:1986ia, Taylor:2003gn}.
 One
can actually consider it as an OSFT counterpart of the high energy
limit of string interactions obtained in \cite{Gross:1987ar,
Moeller:2005ez}. In \cite{Gross:1987ar, Moeller:2005ez} the high
energy limit of string theory is considered in the framework of
perturbative string theory and therefore all string "in" and "out"
states are essentially ''on shell". The vertex we consider
describes off-shell interactions of HS fields as one could expect
from analogy with OSFT. In addition, a remarkable property of our
vertex is that it does not necessarily require the presence of all
kinds of mixed symmetry fields for its classical consistency, in
contrast to OSFT. So one can consider a truncation of the theory
to totally symmetric fields.

In sections \ref{KOm} and \ref{KOmixed} we show how to construct a
cubic vertex for both totally symmetric fields and those with mixed
symmetry. From this vertex we construct the associated nonabelian
deformations of the gauge transformations. We demonstrate how the
algebra of gauge transformations closes to all orders of the
coupling constant $g$. Further on we show that the action is fully
gauge invariant under the nonabelian gauge transformations, which
implies that our vertex is exact. This is analogous to OSFT.

However, one very  important question is to consider a
supersymmetric extension of our construction and in particular for
an HS theory on an $AdS$ background. In this way one can try to make
a connection with the high energy limit of  string theory on a
highly curved Anti de Sitter background, which could be relevant to
AdS/CFT. However, a BRST description of supersymmetric HS theories
on AdS is yet not fully understood. As a simpler problem we consider
the deformation of the proposed bosonic vertex for flat background
to the case of ${\cal D}$ - dimensional AdS. The direct deformation
of the vertex does not seem to be consistent. One can hope that
considering a supersymmetric version of the system will give a
solution result. We leave this for a future investigation.

In any case one can assume that the solution obtained in this paper
might be a step (or at least a nontrivial toy model) towards a
better understanding of the connection between interacting higher
spin gauge theories and string theories.

\setcounter{equation}0\section{Basic equations}\label{BE} A formal
way to construct the nilpotent BRST charge at the high energy limit
is to start with the BRST charge for the open bosonic string
\begin{equation}\label{B1}
{\cal Q} = \sum_{k, l=-\infty}^{+\infty}(C_{-k} L_k - \frac{1}{2}
(k-l):C_{-k}C_{-l}B_{k+l}:)-C_0,
\end{equation}
 perform the rescaling of  oscillator variables
\begin{equation}\label{B2}
c_k = \sqrt{2 \alpha^\prime}C_k, \qquad b_k = \frac{1}{\sqrt{2
\alpha^\prime}}B_k, \qquad c_0 =  \alpha^\prime C_0, \qquad b_0 =
\frac{1}{ \alpha^\prime}B_0,
\end{equation}
$$
\alpha^\mu_k \rightarrow  \sqrt{k} \alpha^\mu_k
$$
and then take  $\alpha^\prime \rightarrow \infty$. In this way one
obtains a BRST charge
\begin{equation}\label{hebrst}
 Q = c_0 l_0 + \tilde{Q} - b_0 {\cal M}
\end{equation}
\begin{equation}\label{B3}
\tilde{Q} = \sum_{k=1}^\infty ( c_k l^{+}_k + c_k^+ l_{k} ), \quad
{\cal M}=\sum_{k=1}^\infty c^{+}_k c_k, \quad l_0 = p^\mu p_\mu,
\qquad l_k^{+}= p^\mu \alpha_{k \mu}^+
\end{equation}
which is nilpotent in any space-time dimension. The oscillator
variables obey the usual (anti)commutator relations
\begin{equation}\label{B4}
[\alpha_\mu^k, \alpha_\nu^{l,+} ] = \delta^{kl} \eta_{\mu \nu},
\quad \{ c^{k,+}, b^l \} = \{ c^k, b^{l,+} \} = \{ c_0^k , b_0^l \}
= \delta^{kl}\,,
\end{equation}
and the vacuum in the Hilbert space is defined as
\begin{equation}\label{B5}
\alpha^\mu_k |0\rangle =  0, \quad
 c_k|0\rangle =  0  \quad k>0 , \qquad
b_k|0\rangle\ = \ 0 \qquad k \geq 0.
\end{equation}
Let us note that one can take the value of $k$ to be any fixed
number without affecting the nilpotency of the BRST charge
(\ref{hebrst}). Fixing the value $k=1$ one obtains the description
of totally symmetric massless higher spin fields, with spins $s,
s-2,..1/0$. The string functional (named ''triplet"
\cite{Francia:2002pt}) in this simplest case has the form
\begin{equation}
\label{Phifield} |\Phi  \rangle = |\phi_1\rangle + c_0
|\phi_2\rangle=  |\varphi \rangle + c^+ \ b^+\ |d\rangle + c_0\
b^+\ |c\rangle \nonumber
\end{equation}

 whereas for an arbitrary value of $k$ one has the so called
''generalized triplet"
\begin{equation}
\label{gentri} |\Phi  \rangle =
 \frac{c^+_{k_1}\dots c^+_{k_p} b^+_{l_1}\dots b^+_{l_p}}{{(p!)}^2}
|D^{l_1, \dots l_p}_{k_1, \dots l_p}\rangle + \frac{c_0
c^+_{k_1}\dots c^+_{k_{p-1}} b^+_{l_1}\dots b^+_{l_p}}{(p-1)! p!}
|C^{l_1, \dots l_p}_{k_1, \dots k_{p-1}}\rangle, \nonumber
\end{equation}
where the vectors $|D^{k_1, \dots k_p}_{l_1, \dots l_p}\rangle $ and
$|C^{k_1, \dots k_p}_{l_1, \dots l_p}\rangle $ are expanded only in
terms of oscillators $\alpha^{\mu +}_k$, and the first term in the
ghost expansion of (\ref{gentri}) with $p=0$ corresponds to the
state $|\varphi \rangle$ in (\ref{Phifield}). One can show that
 the whole spectrum of the open bosonic string decomposes
into an infinite number of generalized triplets, each of them
 describing a finite number of fields with mixed symmetries
\cite{Sagnotti:2003qa}.

In order to describe the cubic interactions  one introduces
 three copies ($i=1,2,3$) of the Hilbert space
defined above, as in bosonic OSFT \cite{Gross:1986ia}. Then the
Lagrangian has the form
\begin{equation} \label {LIBRST}
{L} \ = \ \sum_{i=1}^3 \int d c_0^i \langle \Phi_i |\, Q_i \,
|\Phi_i \rangle \ + g( \int dc_0^1 dc_0^2  dc_0^3 \langle \Phi_1|
\langle \Phi_2|\langle \Phi_3||V \rangle + h.c)\,, \end{equation}
where $|V\rangle$ is the cubic vertex and $g$ is a string
 coupling
constant. The Lagrangian (\ref{LIBRST}) is completely invariant
with respect to the nonabelian gauge transformations
\begin{equation}\label{BRSTIGT1}
\delta | \Phi_i \rangle  =  Q_i | \Lambda_i \rangle  - g \int
dc_0^{i+1} dc_0^{i+2}[(  \langle \Phi_{i+1}|\langle \Lambda_{i+2}|
+\langle \Phi_{i+2}|\langle \Lambda_{i+1}|) |V \rangle] \,,
\end{equation}
provided that the vertex $|V\rangle$ satisfies the BRST invariance
condition
\begin{equation}\label{VBRST}
\sum_i Q_i |V \rangle=0\,.
\end{equation}
The additional constraints imposed by the closure of the algebra of
gauge transformations  will be discussed in section \ref{GKOm}. The
gauge parameter $|\Lambda \rangle$ in each individual Hilbert space
has the ghost structure
\begin{equation}\label{GPdef}
|\Lambda \rangle = b^+ |\lambda\rangle
\end{equation}
 for the totally symmetric case, while the gauge parameters for the generalized triplets
 take the form
\begin{equation}
\label{genGPdef} |\Lambda  \rangle =
 \frac{c^+_{k_1}\dots c^+_{k_p} b^+_{l_1}\dots b^+_{l_{p+1}}}{(p!)(p+1)!}
|\Lambda^{l_1, \dots l_{p+1}}_{k_1, \dots k_p}\rangle + \frac{c_0
c^+_{k_1}\dots c^+_{k_{p-1}} b^+_{l_1}\dots c^+_{l_{p+1}}}{(p-1)!
(p+1)!} |\hat{\Lambda}^{l_1, \dots l_{p+1}}_{k_1, \dots
k_{p-1}}\rangle. \nonumber
\end{equation}
Further on, in order to simplify equations in the following
sections we introduce bilinear combinations of the oscillators
\begin{equation}\label{Defab}
\gamma^{+,ij}_{(kp)}=c^{+,i}_{(k)} b^{+,j}_{(p)}, \quad
\beta^{+,ij}_{(kp)}=c^{+,i}_{(k)} b^j_{0, (p)}\, \quad M^{+,
ij}_{(kp)}= \frac{1}{2} \alpha^{+,\mu,  i}_{(k)} \alpha^{+,j}_{\mu
(p)}
\end{equation}
which have ghost number zero.

Let us make some comments about the BRST charge (\ref{hebrst}). We
can actually justify the way it was obtained from the BRST charge of
the open bosonic string since its cohomologies correctly describe
equations of motion for massless bosonic fields belonging to mixed
symmetry representations of the Poincare group (see e.g.
\cite{Sagnotti:2003qa}). So taking the point of view that, in the
high energy limit the whole spectrum of the bosonic string collapses
to zero mass, which is now infinitely degenerate, one can take the
BRST charge (\ref{hebrst}) as the one which correctly describes this
spectrum.

\setcounter{equation}0\section{An exact cubic vertex for totally
symmetric fields}\label{KOm}

\subsection{BRST invariance}\label{KOmBRST}
We begin first with the simple case of a vertex for totally
symmetric fields. This means we consider only one set of
oscillators as in (\ref{B4}).

The  form of the vertex can be deduced  from the high energy limit
of the corresponding vertex of OSFT. In bosonic OSFT the cubic
vertex has the form
\begin{eqnarray}\label{SFT}
&&| V_3 \rangle = \int \ dp_1\ dp_2\ dp_3\ (2\pi)^d \
\delta^d(p_1+p_2+p_3) \\
&&\times exp \ \left( {1\over 2} \sum_{i,j=1}^3 \
\sum_{n,m=0}^\infty \ \alpha^{+, i}_{n, \mu} \ N^{ij}_{nm}
\alpha^{+, j}_{m, \nu} \ \eta^{\mu \nu} +  \sum_{i,j=1}^3 \
\sum_{n \geq 1, m \geq 0}\ c^{+,i}_{n} X^{ij}_{nm} b_{m}^{+,j}
\right) \ |-\rangle_{123}, \nonumber
\end{eqnarray}
$$
|-\rangle_{123}=c_0^1 c_0^2 c_0^3|0\rangle
$$
where the solution is given in terms of the Neumann coefficients and
all string modes contribute. The oscillators $\alpha^i_{0,\mu}$ are
proportional to the momenta $p^{i}_\mu$. The vertex is  invariant
under the  action of the BRST charge (\ref{B1}). In addition, the
action (\ref{LIBRST})with the vertex (\ref{SFT}) is invariant under
the gauge transformations (\ref{BRSTIGT1})  to all orders in $g$.

As was mentioned in section \ref{BE} at the high energy limit the
BRST charge takes the form (\ref{hebrst}) and can be truncated to
contain any finite number of oscillator variables
\cite{Sagnotti:2003qa}. For this reason it is possible to look for
the BRST invariant vertex that describes the interaction among only
totally symmetric tensor fields of arbitrary rank, without the
inclusion of modes with mixed symmetries.
 One possibility is to start from the SFT vertex
(\ref{SFT}) and keep in the exponential
 only terms proportional to at least one momentum $p^{r}_\mu$,
 therefore dropping all trace operators $(\alpha^r_\mu \eta^{\mu \nu}
\alpha^{s}_\nu)$,
  as one does when obtaining the BRST charge (\ref{hebrst})
  from (\ref{B1}) since
they are leading in the $\alpha ' \to \infty $ limit. However, since
these terms  are exponentiated and  the term $\alpha^{+, r}_{n, \mu}
N^{rs}_{n0} p^{s}_\mu$ is of the same order as $\alpha^{+, r}_{n,
\mu} N^{rs}_{n0} p^{s}_\mu \ (\alpha^{+, r}_{n, \mu} \ N^{rs}_{nm}
\alpha^{+, s}_{m, \nu})^p, \ m,n \geq 1$, a priori one can keep them
both . The same is true  regarding the ghost part where although the
term $c^{+,r}_{n} b_{0}^{s}$ is leading compared to the term
$c^{+,r}_{n} X^{rs}_{nm} b_{m}^{s}, \ n,m \geq 1$ one can not
neglect the later one in the exponential. Let us stress that all
these terms will be essential to maintain the off shell closure of
the algebra of gauge transformations and complete gauge invariance
of the action.

Based on the discussion above one can take the following ansatz for
the vertex which describes interactions between massless totally
symmetric fields with an arbitrary  spin
\begin{equation}\label{KOmod}
|V \rangle=  V^{1} \times V^{mod}|-\rangle_{123}
\end{equation}
where the vertex  contains two parts: a part considered in
\cite{Koh:1986vg}
\begin{equation}\label{KOansatz}
V^{1}= exp\ (\ Y_{ij} l^{+,ij} + Z_{ij} \beta^{+,ij}\ )\,.
\end{equation}
and the part which ensures the closure of the nonabelian algebra
\begin{equation}\label{Vmod}
V^{mod}=exp\ (\ S_{ij} \gamma^{+,ij} + P_{ij} M^{+, ij}\ ),
\end{equation}
where $P_{ij}=P_{ji}$.
 Putting this ansatz into the BRST invariance condition
 one can see that each part of the vertex should be invariant
 separately, namely one obtains
\begin{equation}\label{BRSTD2}
\tilde{ Q} V^{1}|-\rangle_{123}= \sum_i c^{+,i}(Y_{is}
l_0^{is}-Z_{is} l_0^{ss})V^1|-\rangle_{123} =0 \end{equation}
\begin{eqnarray}\label{BRSTD3}
&&\tilde{ Q}\ V^{mod}|-\rangle_{123} =\sum_{i}\ \{ - \
c^{+,i} \ \Bigl({1\over 2} (\delta^{ik}\ l^{+,li} + \delta^{il}\ l^{+,ki})
P_{kl} -  S^{ik} \ l^{+,kk} \Bigr) - \nonumber \\
&&\beta^{+,ii}\ c^{+,m} \ S_{mi} \}V^{mod}|-\rangle_{123} =0
\end{eqnarray}
Using momentum conservation $p_\mu^1+ p_\mu^2 + p_\mu^3=0$ one can
obtain a solution for $Y^{rs}$ and $Z^{rs}$
\begin{equation}\label{KOsolution}
Z_{i,i+1}+Z_{i,i+2}=0
\end{equation}
$$
Y_{i,i+1}= Y_{ii}-Z_{ii} -1/2(Z_{i,i+1}-Z_{i,i+2})
$$
$$
Y_{i,i+2}= Y_{ii}-Z_{ii} +1/2(Z_{i,i+1}-Z_{i,i+2}).
$$
The first term in (\ref{BRSTD3}) should vanish on its own right.
The terms proportional to $c^{+,1}$ are
\begin{eqnarray}\label{BRSTD4}
&&c^{+,1}\ [ P_{12} \ l^{+,21} + P_{13} \ l^{+,31} + P_{11}\
l^{+,11}- S_{12}\ l^{+,22}-S_{13}\ l^{+,33} - S_{11}\ l^{+,11}] \
|-\rangle_{123}=0 \nonumber \\
\end{eqnarray}
with similar terms proportional to $c^{+,2}$ and $c^{+,3}$. Using
momentum conservation to eliminate $p^3_\mu$ we arrive to the
following solution:
\begin{eqnarray}\label{KOmsol}
&& S_{ij}= P_{ij}=0 \qquad i\neq j \\
&& P_{ii} - S_{ii}=0 \qquad i=1,2,3 \nonumber
\end{eqnarray}
It actually turns out that all the other equations resulting from
the ghost expansion of (\ref{BRSTD3}) lead to the same solution. The
second term in (\ref{BRSTD3}) is trivially zero since the off
diagonal components of $S^{ij} \ i \neq j$ vanish. So it does not
result in any additional constraints, but this is not going to be
the case for the mixed symmetry tensor fields considered in the next
section.

\subsection{The gauge transformations and closure of the gauge
algebra }\label{GKOm}

Having determined the form of the vertex from (\ref{KOsolution}) and
(\ref{KOmsol}) we will proceed in computing the commutator of two
gauge transformations with gauge parameters $|\Xi \rangle$ and
$|\Lambda \rangle$. In general, closure of the algebra to order
$O(g)$ implies
\begin{eqnarray}\label{comGT2}
&&[\delta_\Lambda, \ \delta_\Xi] |\Phi_1\rangle =
\delta_{\tilde{\Lambda}} |\Phi_1\rangle= Q_1
|\tilde{\Lambda}_1\rangle -g[( \langle \Phi_2|\langle
\tilde{\Lambda}_3| +\langle \Phi_3|\langle \tilde{\Lambda}_2|) |V
\rangle] + \ O(g^2) \nonumber \\
\end{eqnarray}
where
\begin{equation}\label{GP}
|\tilde{\Lambda}_1 \rangle= g ( \langle \Lambda_2 | \langle
\Xi_3|+\langle \Lambda_3 | \langle \Xi_2|)|V\rangle +\ O(g^2).
\end{equation}
One can check that the closure of the algebra at the first order
in $g$ is equivalent to the BRST invariance of the vertex
\cite{Gross:1986ia} which is satisfied by construction. To ensure
the invariance in higher orders in $g$ one might have to consider
quartic and higher order vertices.
 Obviously it is not guaranteed in general that such a
procedure will lead to the closure of the algebra.
 It should be
emphasized that unlike the case of  free triplets where the total
Lagrangian splits into an infinite sum of individual ones
\cite{Sagnotti:2003qa} for the case of interacting triplets
 the fields
$|\Phi_i\rangle$ of (\ref{Phifield}) need to be an infinite tower of
HS triplet fields, at least when the vertex is defined via
(\ref{KOmod}) or (\ref{KOansatz}). In other words:
\begin{equation}\label{PhiKO}
|\Phi_i\rangle \to \sum_{s=0}^{\infty} \ |\Phi_i^{(s)}\rangle
\end{equation}
In what follows we will assume cyclic symmetry in the three
Hilbert spaces which implies along with (\ref{KOsolution})
\begin{eqnarray}\label{cyclic}
&&Z_{12}=Z_{23}=Z_{31}=Z_a \\
&&Z_{21}=Z_{13}=Z_{32}=Z_b=-Z_a \nonumber \\
&& Y_{12}=Y_{23}=Y_{31}=Y_a \nonumber \\
&&Y_{21}=Y_{13}=Y_{32}=Y_b \nonumber \\
&&Y_{ii}=Y, \ \ Z_{ii}= Z, \ \ P_{ii}=-S_{ii}=-S=P \nonumber
\end{eqnarray}
Finally, it is instructive to give the gauge transformations based
on (\ref{BRSTIGT1})
\begin{eqnarray}\label{GTKOm}
&&\delta (|\varphi_1 \rangle + c_1^+ \ b_1^+\ |d_1\rangle + c^0_1\
b_1^+\ |c_1\rangle)= (l^+_{11}  + l_{11}\ c_1^+ \ b_1^+ + c^0_1\
b_1^+ \ l_0^{1})\ |\lambda_1\rangle  + \nonumber \\
&& + \ g \ e^{(S \ c^+_1  b^+_1)} \ \{- Z_{a}(  \langle
\varphi_2|\ \langle \lambda_3|\ |A\rangle - S\ \left( \langle
d_2|\ \langle
\lambda_3|\ |A\rangle \right)) +  \\
&&+\left(Z\ Z_{a}-Z^2_{b} \right)\ \langle c_2|\ \langle
\lambda_3|\ |A\rangle + \ (2 \leftrightarrow 3, \ Z_a \to Z_b) \}
\nonumber
\end{eqnarray}
where for convenience we have defined the matter part of the
vertex
\begin{equation}\label{A}
|A \rangle= exp\ (\ Y_{ij} l^{+,ij} +P \sum_{i=1}^3 M^{+,ii} )\
|0\rangle.
\end{equation}
In what follows we will show that the vertex defined in
(\ref{KOmod}) allows us to close the algebra of gauge
transformations (\ref{comGT2}) without any order $O(g^2)$
modifications.  Let us assume that there are no quadratic and higher
order terms in $g$ in the gauge transformation law (\ref{comGT2}).
 The commutator of two gauge transformations is
\begin{eqnarray}\label{comGT}
&&[\delta_\Lambda,  \delta_\Xi] |\Phi_1\rangle = Q_1
|\tilde{\Lambda_1}\rangle \\
&&+ g^2  [ \langle V| \left( |\Phi_1\rangle|\Lambda_3\rangle
+|\Lambda_1\rangle|\Phi_3\rangle \ \right) \langle \Xi_3|
|V\rangle+ \langle V| \left( |\Phi_1\rangle|\Lambda_2\rangle +
|\Lambda_1\rangle|\Phi_2\rangle
\right) \langle \Xi_2| |V\rangle  \nonumber \\
&&-\langle V| \left( |\Phi_1\rangle|\Xi_3\rangle +
|\Xi_1\rangle|\Phi_3\rangle \right) \langle \Lambda_3| |V\rangle-
\langle V| \left( |\Phi_1\rangle|\Xi_2\rangle +
|\Xi_1\rangle|\Phi_2\rangle \right) \langle  \Lambda_2| |V\rangle
 \nonumber
\end{eqnarray}
where we have suppressed the integrations over the ghost fields of
(\ref{BRSTIGT1}). Evaluating the LHS of (\ref{comGT}) using
(\ref{BRSTIGT1}) and   plugging it into the RHS of the expression
for $|\tilde{\Lambda}\rangle$ from (\ref{GP})
 we obtain
\begin{eqnarray}\label{comGT3}
&& \langle V| \left( |\Phi_1\rangle|\Lambda_3\rangle +
|\Lambda_1\rangle|\Phi_3\rangle \right) \langle \Xi_3| |V\rangle+
\langle V| \left( |\Phi_1\rangle|\Lambda_2\rangle +
|\Lambda_1\rangle|\Phi_2\rangle
\right) \langle \Xi_2| |V\rangle  \nonumber \\
&&-\langle V| \left( |\Phi_1\rangle|\Xi_3\rangle +
|\Xi_1\rangle|\Phi_3\rangle \right) \langle \Lambda_3| |V\rangle-
\langle V| \left( |\Phi_1\rangle|\Xi_2\rangle +
|\Xi_1\rangle|\Phi_2\rangle \right) \langle  \Lambda_2| |V\rangle
\nonumber \\
&&=\langle V| \left( |\Xi_1\rangle|\Lambda_2\rangle +
|\Lambda_1\rangle|\Xi_2\rangle \right) \langle \Phi_2| |V\rangle +
\langle V| \left( |\Xi_1\rangle|\Lambda_3\rangle +
|\Lambda_1\rangle|\Xi_3\rangle \right) \langle \Phi_3| |V\rangle
\nonumber \\
\end{eqnarray}
Equation (\ref{comGT3}) is valid for any value of the vertex, but
using the  solution  (\ref{KOmod}), (\ref{KOsolution}),
(\ref{KOmsol}) one can see that
  the expression in (\ref{GP}) vanishes identically.
 A typical term to demonstrate this is:
\begin{eqnarray}\label{GPKOm}
&&\langle \tilde{\Lambda}_3| \sim \langle V | |\Xi_1 \rangle
|\Lambda_2\rangle= \int dc_0^1 \
dc_0^2\ _{123}\langle-| exp\ (-Z_{ij}\ c^i b_0^j - S^*\  c^i b^i) \nonumber \\
&&\times \  b^+_1\ b^+_2 (\langle A| |\xi_1\rangle |\lambda_2\rangle
|0_1\rangle_{gh} |0_2\rangle_{gh}.
\end{eqnarray}
It is impossible to make the RHS of the expression above
proportional to $b^3$ as  is necessary from (\ref{GPdef}) since this
requires the vertex to have terms like $c^2b^3$ and $c^1b^3$ which
are absent since $S_{ij}$ is diagonal. Now this implies that the RHS
of (\ref{comGT3}) vanishes identically. Therefore the LHS of
(\ref{comGT3}) will have to vanish. Actually it will turn out that
each term of the LHS of (\ref{comGT3}) vanishes identically. Since
the computation is  long  we present only a representative term:
\begin{eqnarray}\label{comGT4}
&&\langle V|  |\Phi_1\rangle|\Lambda_3\rangle \ \langle \Xi_3|
|V\rangle= \int dc_0^2 \ dc_0^3\ \ ( \int dc_0^3 \ dc_0^1\
_{123}\langle-| \ V^{\dagger, gh} \ \langle A^\dagger| \times \\
&& ( |\varphi_1\rangle + c^+_1 b^+_1 |d_1\rangle + c_0^1 b^+_1
|c_1\rangle) b^+_3 |\lambda_3\rangle ) \times \nonumber \\
&&\times \langle \xi_3| b_3 |A\rangle \ V^{gh} |-\rangle_{123}
\nonumber
\end{eqnarray}
where we have denoted
\begin{equation}\label{Vgh}
V^{gh}= exp\ \left( Z_{ij}\ c^i b_0^j + S\  c^i b^i \right)
\end{equation}
The integration in the parentheses can be easily performed and it
gives us a state in Hilbert space number two
\begin{equation}\label{comGT5}
- \ _{gh}\langle 0_2| \left( Z_b\ (\langle
A^\dagger|\varphi_1\rangle-S\langle A^\dagger|d_1\rangle)- (Z\
Z_b- Z^2_a) \langle A^\dagger|c_1\rangle \right) |\lambda_3
\rangle exp\ (-S^*\ c^2 b^2)
\end{equation}
Inserting the expression above in (\ref{comGT4}) and performing
the second integration of ghosts we get
\begin{eqnarray}\label{comGT6}
&& Z_a  \{Z_b(\langle A^\dagger|\varphi_1\rangle-S\langle
A^\dagger|d_1\rangle)- (Z\ Z_b- Z^2_a) \langle
A^\dagger|c_1\rangle \}|\lambda_3 \rangle \ \times \nonumber
\\
&&\times (1-|S|^2) \ \langle \xi_3| A\rangle \ exp\ (S c^+_1
b^+_1) |0_1\rangle_{gh}
\end{eqnarray}
where the factor $(1-|S|^2)$ came from saturating the second
Hilbert space ghost vacuum:
\begin{eqnarray}\label{comGT7}
&&\int dc_0^2\ _{gh}\langle\ 0_2|\ exp\ (-S\ c^2 b^2) \ V^{gh} \
|-\rangle_{123}= \\
&&-(1-S^2) exp\ (Z_{ij}\ c^i b_0^j + S\  c^i b^i)|_{i,j \neq 2} \
c_0^1\ c_0^3 |0_1\rangle_{gh}|0_3\rangle_{gh}.\nonumber
\end{eqnarray}
Proceeding in the same manner we can compute all terms and collect
them in (\ref{comGT3}). The expressions proportional to $|\varphi
\rangle$ and $|d\rangle$ in (\ref{comGT6})  cancel among each other
without any further constraints on the parameters. On the other
hand, cancelation of terms proportional to $|c\rangle$ in
(\ref{comGT6}) is non trivial. One can check a few
 examples which involve lower numbers of derivatives in the
interaction part of the Lagrangian.

Checking the invariance  with respect to the gauge transformations
which  involve  the gauge parameters $\lambda$, $\xi$
 and the scalar field
$c$ one can suppose the existence of a solution
  $Y=Z=0$ and
$Y_a=-Z_a$, $Y_b=Z_a$ with an arbitrary value of $S$. However, this solution does not allow one to
close the algebra when  the parameters of gauge transformations of
 a higher rank  tensor states $c_{\mu_1 \dots
\mu_n}$ are involved.
 Therefore (\ref{comGT6}) imposes \footnote{ Note
that there is a potential infinity coming from the matter part of
(\ref{comGT6}). Indeed there there terms $\langle 0| e^{(P M
)} \ e^{(P  M^+)} |0\rangle$. Depending on the
dimensionality of the space-time ${\cal D}$ it can have a
logarithmic divergence or higher. This might seem to invalidate
our condition (\ref{clGT}), but nevertheless, the condition
(\ref{clGT}) makes the expression in (\ref{comGT6}) strictly zero.
 So  in the spirit  of (\ref{dphi3}) we have an infinite series of
contributions which cancel term by term. We believe that this is
the correct prescription in dealing with this issue.}
\begin{equation}\label{clGT}
|S|^2=1 \ .
\end{equation}
Actually (\ref{clGT}) implies that each term of (\ref{comGT3})
should vanish separately.  This leads to a  trivial commutator:
\begin{equation}\label{comGT8}
[\delta_\Lambda,  \delta_\Xi] |\Phi_1\rangle =0
\end{equation}
or rather to
\begin{equation}\label{comGT9}
\delta_\Lambda \delta_\Xi | \Phi_1 \rangle=0.
\end{equation}
In other words we can consider the vertex (\ref{KOmod}) as a field
dependent deformation of the  BRST charge in (\ref{hebrst}), which
can be written schematically
\begin{equation}\label{newbrst}
Q'= Q + gV(\Phi)
\end{equation}
with the nilpotency property
\begin{equation}\label{nilbrst}
Q'^2= Q^2 + 2g Q V(\Phi) + g^2 V(\Phi)^2=0
\end{equation}
which follows from the nilpotency of $Q$,  the BRST invariance of
the vertex (\ref{VBRST}) and  (\ref{clGT}). Proceeding further with
analogy with the String Field Theory one can make both the string
functional and gauge transformation parameters to be matrix valued
(i.e., introduce Chan --Paton factors). The resulting theory will
still satisfy (\ref{comGT9}).

Now we should check whether the cubic vertex we constructed is
exact. In other words the action (\ref{LIBRST}) should be invariant
under (\ref{BRSTIGT1}) to all orders in $g$. We should point out
that closure of the algebra does not ensure this. The simplest
counter example is the $U(1)$ scalar electrodynamics. The gauge
transformation of the scalar, $\delta \varphi \sim i \lambda
\varphi$ is derived from  (\ref{BRSTIGT1}) and the form of the cubic
interaction of scalar electrodynamics. The algebra of gauge
transformations is abelian as in (\ref{comGT8}). Nevertheless, the
action is not fully invariant and requires a quartic term.

To demonstrate the invariance of (\ref{LIBRST}) in our case it is
instructive to consider an example. From (\ref{LIBRST}) we can
easily deduce the presence of the term with no derivatives
\begin{equation}\label{phi3}
V_{\varphi \varphi \varphi}= -g \ \left( \varphi -S\ d- Z c
\right)^3 \ + \ h.c  + \ \dots
\end{equation}
where $\varphi$, $c$ and $d$ are the scalar components of
(\ref{PhiKO}) from triplets with spin $0,1,2$ respectively. The
omitted terms are proportional to $c^2, c^3$ and have a gauge
variation which does not mix with the gauge variation of the term
in (\ref{phi3}), so they will not affect our arguments bellow. The
gauge variation of this term with respect to the scalar gauge
parameter $\lambda$ results in $O(g^2)$ terms \footnote{We have
assumed that the coefficients $S_{ii}$ and $Z_{ij}$ are in general
complex}
\begin{equation}\label{dphi3}
\delta V_{\varphi \varphi \varphi}= +3g^2\ (1-|S|^2) \ [ (Z \ Z_a
-Z_b^2) + (Z_b\ Z-Z_a^2)] \ c\ \lambda \left( \varphi -S\ d- Z c
\right)^2.
\end{equation}
This variation cannot be cancelled against the gauge variation of
any other cubic term since the latter contain derivatives. The
vanishing of (\ref{dphi3}) after using $Z_b=-Z_a$ requires
$|S|^2=1$. We see therefore from (\ref{GTKOm}) that invariance of
the action requires the non-abelian part of the gauge
transformation of the scalar $\varphi$ to cancel that of the
scalar $d$ which is the $s=0$ field of the $s=2$ triplet. This is
suggests that the whole tower of HS fields is necessary since the
gauge transformation of each "top-spin" component of each triplet
will be canceled against the gauge transformation of the
"bottom-component' of another triplet. Notice that just as  we
explained below (\ref{comGT7}) the
 condition
(\ref{clGT}) is imposed in order to cancel terms $\sim \ |c\rangle
|\lambda \rangle$ when one checks  the closure of the algebra. It
is easy to deduce that the full action is invariant under
(\ref{BRSTIGT1}) with the vertex (\ref{KOmod}).

Let us make several comments:

\begin{itemize}
\item Dropping the cyclicity  constraint does not seem to alter the
conclusions. In this case we will have $|S_{ii}|^2=1, \ i=1,2,3$.

\item Despite the algebra being trivial it seems the vertex cannot
be obtained from the free Lagrangian via some field redefinition. In
other words the vertex (\ref{KOmod}) is not trivial in cohomologies
of the BRST charge (\ref{hebrst}): $|V\rangle \neq Q|W\rangle$ for
some $|W \rangle$. One can show that only terms diagonal in the
Hilbert spaces $i,j$ can be removed from the exponent of
(\ref{KOmod}) via a specific field redefinition scheme
\cite{Buchbinder:2006eq}. Moreover, notice that in (\ref{phi3}) the
presence of both scalars $\varphi$ and $d$ is required in order to
have gauge invariance. The nonabelian gauge transformation of the
one cancels the nonabelian part of the other. Their abelian parts
differ though: The scalar $\varphi$ has trivial abelian gauge
transformation while $\delta d\sim
\partial^\mu \lambda_\mu$ as  required by the gauge invariance
of the spin two triplet. If one tries to remove the vertex by a
field redefinition from  the free Lagrangian one will mix
  the free equations of the triplets, which as we know decouple in
  the free limit.

\item The infinite tower of triplets is essential for the closure. The
nonabelian part of the gauge transformation of each component of
$|\varphi\rangle$ is canceled against the same rank tensor component
of $|d \rangle$. However, the two tensors belong though to different
triplets.

\item In order to deal with the quantum theory one needs to gauge
fix the action. The usual partial gauge fixing condition is the
Siegel gauge $b_0 |\phi \rangle =0 $. This gauge  eliminates
auxiliary fields $|c\rangle$ as one can see from (\ref{Phifield}).
Then obviously the algebra on the constrained string field closes
without any constraint on the parameter $S$. However, the off-shell
closure of the algebra  requires (\ref{clGT}).

\end{itemize}

\setcounter{equation}0\section{An exact cubic vertex for mixed
symmetry fields}\label{KOmixed}

The case of arbitrary mixed symmetry fields is completely analogous
to the construction in section \ref{KOm} for totally symmetric
fields. As in (\ref{KOmod}) we make the ansatz
\begin{eqnarray}\label{KOmmod}
&&V= exp\ (\sum_{n=1}^\infty \ Y^{(n)}_{ij} l_{ij}^{+,(n)} +
 \ Z^{(n)}_{ij} \beta_{ij}^{+,(n)})\times  \\
&&exp\ (\sum_{n,m=1}^\infty\ S^{(nm)}_{ij} \gamma^{+,(nm)}_{ij} +
P^{(nm)}_{ij} M^{+,(nm)}_{ij}) \nonumber
\end{eqnarray}
where in this case we are summing over $n,m$  as well. We put the
oscillator level indices in parentheses in order to distinguish them
from the Hilbert space ones. The oscillator algebra takes the form
\begin{equation}\label{oscialg}
[\alpha_\mu^{(m),i}, \alpha_\nu^{+,(n),j} ] =
\delta^{mn}\delta^{ij} \eta_{\mu \nu}, \quad \{ c^{+,(m),i},
b^{(n),j} \} = \{ c^{(m),i}, b^{+,(n),j} \} =
\delta^{mn}\delta^{ij}.
\end{equation}
The BRST invariance with respect to (\ref{hebrst}) implies
(\ref{BRSTD2}) and (\ref{BRSTD3})
\begin{eqnarray}\label{BRSTM1}
&&\sum_{i=1}^3 \ \sum_{r=0}^\infty \ c^{+,(r),i} ( Y^{(r)}_{is}
l_0^{is}-Z^{(r)}_{is} l_0^{ss})|-\rangle_{123} =0
\end{eqnarray}
\begin{eqnarray}\label{BRSTM2}
&&\sum_{i=1}^3\ \sum_{r=0}^\infty \{  \ c^{+,(r),i} \
\Bigl( {1\over 2}
(P_{il}^{(rs)}\ l^{+,(s),li} + P_{li}^{(sr)}\ l^{+,(s),li}) -  S^{(rs)}_{ik} \ l^{+,(s),kk} \Bigr)  - \nonumber \\
&& b^i_0 \ c^{+,(p),i}\ c^{+,(r),m} \ p\  S^{(rp)}_{mi}
\}|-\rangle_{123} =0
\end{eqnarray}
where the summation over repeated indexes is assumed. Solving
(\ref{BRSTM1}) we get
\begin{equation}\label{KOMsolution}
Z^{(r)}_{i,i+1}+Z^{(r)}_{i,i+2}=0
\end{equation}
\begin{equation}
Y^{(r)}_{i,i+1}= Y^{(r)}_{ii}-Z^{(r)}_{ii} -{1 \over
2}(Z^{(r)}_{i,i+1}-Z^{(r)}_{i,i+2})
\end{equation}
\begin{equation}
Y^{(r)}_{i,i+2}= Y^{(r)}_{ii}-Z^{(r)}_{ii} +{1\over
2}(Z^{(r)}_{i,i+1}-Z^{(r)}_{i,i+2}).
\end{equation}
Equation (\ref{BRSTM2}) gives
\begin{eqnarray}\label{KOMmsol}
&& S^{(ps)}_{ij}= P^{(ps)}_{ij}=0 \qquad i\neq j \ or \ p \neq s \\
&&  P^{(ss)}_{ii} - S^{(ss)}_{ii}=0 \qquad i=1,2,3 \nonumber
\end{eqnarray}
where  unlike the case of  totally symmetric fields the second term
of (\ref{BRSTM2}) is not identically zero before taking matrices $S$
and $P$ to be diagonal in $p$ and $s$ . We can once more choose a
cyclic solution in the three Hilbert spaces as in (\ref{cyclic}) and
in this way get  an obvious generalization of (\ref{KOMsolution}).

  The discussion of the closure of the algebra follows closely the
lines of  subsection \ref{GKOm}. The condition for the closure  is
again the equation (\ref{comGT3}). In addition,  in the case of both
mixed symmetry and  totally symmetric fields
 we have a diagonal $S^{(ps)}_{ij}$. It is straightforward to show
that $|\tilde{\Lambda}\rangle$ vanishes as  in (\ref{GPKOm}). The
steps similar to (\ref{comGT4}) and (\ref{comGT5}) lead us to the
equivalent of (\ref{comGT6})
\begin{eqnarray}\label{comMGT1}
&&\langle V|  |\Phi_1\rangle|\Lambda_3\rangle \ \langle \Xi_3|
|V\rangle=    \sum_{n=0}^\infty\ Z^{(r_{n+1}}_a {\cal A} \ _3
\langle \Xi^{r_1 \dots r_n)}_{r_1 \dots r_n}|A\rangle  \ T^{(r_1
\dots
r_{n+1})} \\
&& \times \left(\sum_{m=0}^\infty  \prod_{r \in{\cal
S}_m}^m(1-|S^{(r)}|^2)\ \right) e^{(\sum_p\ S^{(p)} c^{+,(p)}_{1}
b^{+,(p)}_{1})}|0\rangle_{123} \nonumber
\end{eqnarray}
where we have shown only the terms proportional to the gauge
parameter $|\Lambda^{j_1 \dots j_{l+1}}_{i_1 \dots i_l} \rangle$
\begin{eqnarray}\label{AA}
&&{\cal A} \ \sim \ -\langle 0_2| \sum_{l=0}^{\infty}\ \left(
\langle A|D^{(i_1, \dots i_l)}_{i_1, \dots i_l}\rangle_1 \
|\Lambda^{(j_1, \dots j_{n+1})}_{j_1, \dots j_n}\rangle_3 \right)
\ \sum_{j_{n+1}} Z_b^{(j_{n+1}} \ T_1^{(i_1
\dots i_l)}\ T_3^{j_1 \dots j_n)} + \nonumber \\
 &&+\langle 0_2| \
c_0^2 \ \left( \langle A|C^{(i_1, \dots i_{l+1})}_{i_1, \dots
i_l}\rangle_1 \ |\Lambda^{(j_1, \dots
j_{n+1})}_{j_1, \dots j_n}\rangle_3 \right)\  \nonumber \\
&&\times \sum_{i_{l+1}, j_{n+1}} \ (- Z^{(i_{l+1}} Z_b^{(j_{n+1}}
+ Z_a^{(i_{l+1}} Z_a^{(j_{n+1}}) \ T_1^{i_1 \dots i_l)} \ T_3^{j_1
\dots j_n)}
\end{eqnarray}
and the set ${\cal S}_m$ is defined as the set of all partitions of
$\mathbb{Z}$ in subsets of $m$ arbitrary integers. Obviously this is
an infinite set. The indices $(i_r, j_r)$ are assumed to take values
in $\mathbb{Z}$ and label the levels of the oscillators involved.
Every term in the summations over $l$ in (\ref{AA}) or $n$ in
(\ref{comMGT1}) has an infinite number of terms. This comes about
since there are infinite tensor states of rank $l$(or rank $n$).
These are labeled from the infinite number of subsets in
$\mathbb{Z}$ which are made by $l$ (or $m$) random integers.
Obviously, in tension-full string theory these tensors have
different masses, while in the tensionless limit all of them become
massless. The tensor $T_i^{j_1 \dots j_n}$ is defined as
\begin{equation}\label{T}
T_i^{j_1 \dots j_n}= \sum_{l=0}^\infty \ \sum_{perm \ (j_r)} \
\left( \prod_{r=1}^l \ S_i^{(j_r)} \right) \ (-1)^{P(j_1 \dots
j_l)} (-1)^l
\end{equation}
where $P( j_1 \dots j_l)$ stands for the permutations of the
indices $(j_1 \dots j_l )$. Then $(-1)^{P(j_1 \dots j_l)}$ gives
$+1$ for any even permutation of the set $(j_1 \dots j_l)$ and
$-1$ for an odd permutation. The tensor defined above effectively
sums over all possible ways of saturating the ghosts of vertex
(\ref{KOmmod})  with a given state like (\ref{gentri}) or
(\ref{genGPdef}). In (\ref{AA}) the indices in parentheses are
assumed to be contracted and therefore summed over. In a state
$|\Lambda^{(j_1, \dots j_{n+1})}_{j_1, \dots j_n}\rangle$ it is
implied that the the first $n-$upper indices are equal to the
lower ones and only the last $j_{n+1}$ can differ.

Our expressions  (\ref{comMGT1}) and (\ref{AA}) are  divergent due
to the infinite degeneracy of the tensionless string spectrum.
Nevertheless, as in the case of the totally symmetric states, one
can show that the first term in (\ref{AA}) can cancel among the four
different contributions on the LHS of (\ref{comGT3}). The second
term though can cancel only if
\begin{equation}\label{clMGT}
|S^{(r)}|\ ^2=1, \qquad  \forall \ r \ \in \mathbb{Z}.
\end{equation}
The condition above can be shown along the lines of section
\ref{GKOm} to ensure invariance of the action for the case of mixed
symmetry fields as well.

This completes our treatment of the the whole spectrum of the open
bosonic string at the high energy limit. The mechanism of the
closure of the algebra and consequent gauge invariance of the vertex
is the same as that in the ubsection \ref{GKOm}. Again,  in the
Siegel gauge (if imposed)  the  algebra of gauge transformations
with constrained parameters closes without any constraint on the
coefficients $S^{(r)}_{ii}$ due to non-trivial cancelation among the
four surviving terms of the LHS of (\ref{comGT3}).

\setcounter{equation}0\ \section{AdS space}\label{ADSm} In order to
extend the discussion of the previous section to the case of an
arbitrary dimensional $AdS$ space let us recall some relevant facts
about the triplet formulation on anti --de Sitter space
\cite{Sagnotti:2003qa}-- \cite{Fotopoulos:2006ci} (see also
\cite{Buchbinder:2001bs}-- \cite{Buchbinder:2006ge}) We restrict
ourselves  to the case of totally symmetric fields on ${\cal
D}$-dimensional AdS space-time i.e., to the case of the ''triplet".
 The formulas given in
 section \ref{BE} apply to the case of AdS space as well (see
\cite{Buchbinder:2006eq} for details) , but now the ordinary
partial derivative replaced by the operator
\begin{equation}
\label{pop} p_\mu \ = \  -\; i \, \left( \nabla_\mu +
\omega_{\mu}^{ab} \, \alpha_{\; a}^+\,
  \alpha_{ \; b} \right) \ ,
\end{equation}
where    $\omega_\mu^{ab}$ is  the spin connection of AdS and
$\nabla_\mu$ is the AdS covariant derivative.  The AdS counterpart
of the BRST charge (\ref{hebrst}) has the form
\begin{eqnarray} \label{brst}
Q& =&c_0( l_0 \, + \, \frac{1}{L^2} (  N^2 - 6N +6 +{\cal D} \, -
\, \frac{{\cal D}^2}{4} - 4M^+M +
 c^+b(4N-6)  \\ \nonumber
&+& b^+c (4N-6) + 12 c^+b b^+c -8 c^+b^+M  +8 M^+bc)) + c^+l +
cl^+ - c^+cb_0
\end{eqnarray}
where $l_0$ is the AdS covariant d'Alembertian, $L$ is the radius of
the AdS space and
\begin{equation} N \ = \ \alpha^{\mu +}  \alpha_{\mu} \ + \
\frac{\cal D}{2} , \quad M= \frac{1}{2} \alpha^\mu \alpha_\mu \,.
\label{Nop}
\end{equation}
Following the same strategy as in the case of flat space time one
can try to make an AdS deformation of the flat space--time solution.
However, a one can find that the direct AdS deformation of the
solution (\ref{KOmod}) does not exist. In other words one can not
find appropriate ''counterterms" proportional to $\frac{1}{L^2}$  in
the vertex (\ref{KOmod}) which would make it a solution of
equation\footnote{It does not mean of course that this equation does
not have a solution at all \cite{Irges}.} (\ref{VBRST}).

The simplest way to see the problem is as follows. First let us
note \cite{Buchbinder:2006eq} that the operator
\begin{equation}\label{DefN}
\tilde N=\alpha^{\mu, i+} \alpha^i_{\mu}+b^{i,+} c^{i}+
c^{i,+}b^i\,
\end{equation}
commutes with the BRST charge (\ref{brst}). So it is sufficient to
check the consistency of (\ref{VBRST}) to any given level
(eigenvalue of $\tilde{N}$). Expanding vertex  (\ref{KOmod}) to the
first level it is straightforward to show that BRST invariance of
the vertex is maintained. Expanding to level two we get the
following terms:
\begin{eqnarray}\label{level2}
&&V  \ \sim \ {1\over 2} Y_{ij} Y_{mn} l^{+,ij} l^{+,mn} + Y_{ij}
Z_{mn} l^{+,ij} \beta^{+,mn} + \nonumber \\
&& + {1\over 2} Z_{ij} Z_{mn} \beta^{+,ij}\beta^{+,mn} + P_{ij}
M^{+,ij} + S^{ij} \gamma^{+,ij}.
\end{eqnarray}
The coefficients $Z_{ij}$, $Y_{ij}$, $P_{ij}$ and $S_{ij}$ are now
general functions of the AdS radius $L^2$. They are assumed to have
an ${1\over L^2}$ expansion with the zeroth order term given by
(\ref{KOsolution}) and (\ref{KOmsol}). One can easily see that just
as in the flat case the last term of (\ref{brst}) results in
$S_{ij}$ to be diagonal. Further on the ${1\over L^2}$ terms of
$\hat{l_0}$ result in terms like $c^{+,i} \alpha^{+,i} p^{j}, \ j
\neq i$. These terms can only be canceled if we set $Y_{12}=Y_{13}$.
That implies $Z_a=0$ and all off-diagonal components of the vertex
are zero. This is a trivial vertex. So there is no non-trivial
solution for AdS of the form (\ref{KOmod}).

A possible heuristic explanation of this fact is that anti--de
Sitter space is not a solution of bosonic string theory. One might
try to generalize the procedure described above to the high energy
limit of the open superstring. This would require an analogous
nilpotent BRST charge for fermionic massless higher spin fields as
well as for mixed symmetry fields,
 thus finding a Lagrangian description for the
equations obtained in \cite{Metsaev:1997nj} on possibly
$AdS_{{\cal D}}\otimes S_{{\cal D^\prime}}$ background
 along the lines of
\cite{Sagnotti:2003qa}-- \cite{Fotopoulos:2006ci}. This seems
possible since the BRST charge (\ref{brst})is nilpotent not only
for the case of Anti -- de Sitter space but for any space--time
with a constant curvature as well. In this case an AdS deformation
of the vertex (\ref{KOmod}) might exist. We leave this interesting
question for future work.

 \vspace{1cm}

\noindent {\bf Acknowledgments.} It is a pleasure to thank X.
Bekaert, N. Boulanger, I. L. Buchbinder, P. Cook, C. Iazeolla, N. Irges, A. Petkou,
S. Sciuto, P. Sundell, A. Sagnotti and P. West for valuable
discussions. M.T. would like to thank Scuola Normale Superiore
(Pisa, Italy) and Department of Physics of Torino University
(Turin, Italy), where a part of this work has been done. The work
of A.F. is partially supported by the European Commission, under
RTN program MRTN-CT-2004-0051004 and by the Italian MIUR under the
contract PRIN 2005023102. The work of M.T. was supported by the
European Commission Under RTN program MRTN-CT-2004-512194.


\renewcommand{\thesection}{A}

\setcounter{equation}{0}

\renewcommand{\theequation}{A.\arabic{equation}}


\begin{thebibliography}{99}

\bibitem{Gross:1987ar}
   D.~J.~Gross and P.~F.~Mende,
   Nucl.\ Phys.\ B {\bf 303}, 407 (1988).


\bibitem{Moeller:2005ez}
   N.~Moeller and P.~West,
   Nucl.\ Phys.\ B {\bf 729} (2005) 1
   [arXiv:hep-th/0507152].

\bibitem{Bandos:2007qn}
  I.~A.~Bandos, J.~A.~de Azcarraga and C.~Miquel-Espanya,
  [arXiv:hep-th/0702133]. \\
  P.~Horava and C.~A.~Keeler,
  arXiv:0704.2230 [hep-th].

\bibitem{Sezgin:2002rt}
   E.~Sezgin and P.~Sundell,
   Nucl.\ Phys.\ B {\bf 644}, 303 (2002)
   [Erratum-ibid.\ B {\bf 660}, 403 (2003)]
   [arXiv:hep-th/0205131].\\
   M.~Bianchi, J.~F.~Morales and H.~Samtleben,
   JHEP {\bf 0307}, 062 (2003)
   [arXiv:hep-th/0305052]. \\
  P.~Haggi-Mani and B.~Sundborg,
  JHEP {\bf 0004} (2000) 031
  [arXiv:hep-th/0002189].

\bibitem{Vasiliev:1990en}
   M.~A.~Vasiliev,
   Phys.\ Lett.\ B {\bf 243}, 378 (1990). \\
   M.~A.~Vasiliev,
   Class.\ Quant.\ Grav.\  {\bf 8}, 1387 (1991). \\
   M.~A.~Vasiliev,
   Phys.\ Lett.\ B {\bf 567}, 139 (2003)
   [arXiv:hep-th/0304049]. \\
  M.~A.~Vasiliev,
  Fortsch.\ Phys.\  {\bf 52} (2004) 702
  [arXiv:hep-th/0401177].

\bibitem{Francia:2002pt}
   D.~Francia and A.~Sagnotti,
   Class.\ Quant.\ Grav.\  {\bf 20} (2003) S473
   [arXiv:hep-th/0212185].


\bibitem{Sagnotti:2003qa}
   A.~Sagnotti and M.~Tsulaia,
   Nucl.\ Phys.\ B {\bf 682} (2004) 83
   [arXiv:hep-th/0311257].

\bibitem{Francia:2006hp}
  D.~Francia and A.~Sagnotti,
  J.\ Phys.\ Conf.\ Ser.\  {\bf 33}, 57 (2006)
  [arXiv:hep-th/0601199].


\bibitem{Fotopoulos:2006ci}
  A.~Fotopoulos, K.~L.~Panigrahi and M.~Tsulaia,
  Phys.\ Rev.\  D {\bf 74} (2006) 085029
  [arXiv:hep-th/0607248].



\bibitem{Buchbinder:2006eq}
  I.~L.~Buchbinder, A.~Fotopoulos, A.~C.~Petkou and M.~Tsulaia,
  Phys.\ Rev.\  D {\bf 74} (2006) 105018
  [arXiv:hep-th/0609082].

\bibitem{Buchbinder:2001bs}
   I.~L.~Buchbinder, A.~Pashnev and M.~Tsulaia,
   Phys.\ Lett.\ B {\bf 523} (2001) 338
   [arXiv:hep-th/0109067]. \\
   I.~L.~Buchbinder, A.~Pashnev and M.~Tsulaia,
   [arXiv:hep-th/0206026].\\
  A.~Pashnev and M.~Tsulaia,
  Mod.\ Phys.\ Lett.\  A {\bf 13}, 1853 (1998)
  [arXiv:hep-th/9803207].



\bibitem{Buchbinder:2007ak}
  I.~L.~Buchbinder, A.~V.~Galajinsky and V.~A.~Krykhtin,
  [arXiv:hep-th/0702161]. \\
  I.~L.~Buchbinder, V.~A.~Krykhtin and A.~A.~Reshetnyak,
  [arXiv:hep-th/0703049].


\bibitem{Buchbinder:2006ge}
   I.~L.~Buchbinder, V.~A.~Krykhtin and P.~M.~Lavrov,
   [arXiv:hep-th/0608005].



\bibitem{Koh:1986vg}
   I.~G.~Koh and S.~Ouvry,
   Phys.\ Lett.\ B {\bf 179}, 115 (1986)
   [Erratum-ibid.\  {\bf 183B}, 434(E) (1987)]
\bibitem{Bonelli:2003kh}
   G.~Bonelli,
   Nucl.\ Phys.\ B {\bf 669}, 159 (2003)
   [arXiv:hep-th/0305155].





\bibitem{Bekaert:2005jf}
   X.~Bekaert, N.~Boulanger and S.~Cnockaert,
   JHEP {\bf 0601}, 052 (2006)
   [arXiv:hep-th/0508048].\\
N.~Boulanger, S.~Leclercq and S.~Cnockaert,
Phys.\ Rev.\ D {\bf 73} (2006) 065019 [arXiv:hep-th/0509118]. \\
   R.~R.~Metsaev,
  [arXiv:hep-th/0512342]. \\
  E.~Sezgin and P.~Sundell,
  Nucl.\ Phys.\  B {\bf 762} (2007) 1
  [arXiv:hep-th/0508158].
  P.~Benincasa and F.~Cachazo,
  arXiv:0705.4305 [hep-th].



\bibitem{Witten:1985cc}
  E.~Witten,
  Nucl.\ Phys.\  B {\bf 268}, 253 (1986).



\bibitem{Gross:1986ia}
   D.~J.~Gross and A.~Jevicki,
   Nucl.\ Phys.\ B {\bf 283}, 1 (1987). \\
   D.~J.~Gross and A.~Jevicki,
   Nucl.\ Phys.\ B {\bf 287}, 225 (1987).\\
   A.~Neveu and P.~C.~West,
   Nucl.\ Phys.\ B {\bf 278} (1986) 601.


\bibitem{Taylor:2003gn}
  W.~Taylor and B.~Zwiebach,
  [arXiv:hep-th/0311017].

\bibitem{Irges}
A.~ Fotopoulos, N.~ Irges, A.~C.~Petkou and M.~Tsulaia (to appear)

\bibitem{Metsaev:1997nj}
   R.~R.~Metsaev,
   [arXiv:hep-th/9810231].\\
  R.~R.~Metsaev,
  Phys.\ Lett.\  B {\bf 419} (1998) 49
  [arXiv:hep-th/9802097].




\end{thebibliography}
\end{document}